# Magnetization and anomalous Hall effect in SiO$_2$/Fe/SiO$_2$ trilayers

Sudhansu Sekhar Das and M. Senthil Kumar[1]

*Department of Physics, Indian Institute of Technology Bombay, Mumbai 400 076, India*

**Abstract:** SiO$_2$/Fe/SiO$_2$ sandwich structure films fabricated by sputtering were studied by varying the Fe layer thickness ($t_{Fe}$). The structural and microstructural studies on the samples showed that the Fe layer has grown in nanocrystalline form with (110) texture and that the two SiO$_2$ layers are amorphous. Magnetic measurements performed with the applied field in in-plane and perpendicular direction to the film plane confirmed that the samples are soft ferromagnetic having strong in-plane magnetic anisotropy. The temperature dependence of magnetization shows complex behavior with the coexistence of both ferromagnetic and superparamagnetic properties. The transport properties of the samples as studied through Hall effect measurements show anomalous Hall effect (AHE). An enhancement of about 14 times in the saturation anomalous Hall resistance ($R_{hs}^A$) was observed upon reducing the $t_{Fe}$ from 300 to 50 Å. The maximum value of $R_{hs}^A$ = 2.3 Ω observed for $t_{Fe}$ = 50 Å sample is about 4 orders of magnitude larger than that reported for bulk Fe. When compared with the single Fe film, a maximum increase of about 56% in the $R_{hs}^A$ was observed in sandwiched Fe (50 Å) film. Scaling law suggests that the $R_s$ follows the longitudinal resistivity ($\rho$) as, $R_s \propto \rho^{1.9}$, suggesting side jump as the dominant mechanism of the AHE. A maximum enhancement of about 156% in the sensitivity S was observed.

---

[1]Corresponding Author E-mail: senthil@iitb.ac.in

# 1. Introduction

Recently, ferromagnetic heterostructures are being studied extensively because of their interesting magnetotransport properties, such as anisotropic magnetoresistance, giant magnetoresistance and tunnel magnetoresistance. Composite metal-insulator systems have brought the attention of the researchers because of their complex electrical and magnetotransport properties [1-3]. Anomalous Hall effect (AHE) which is also a spin dependent transport phenomenon is mostly observed in magnetic materials [4]. The relatively smaller values of the AHE in bulk magnetic materials have limited its application possibilities in the field of spintronics. Linear field response, hysteresis free behavior with field and low manufacturing cost of the AHE devices pioneers new avenues for future exploration of the effect. Recently, some groups have also reported that the AHE being magnetization dependent could be used as a tool for magnetic characterization of nanostructures, thin films, recording layer of double-layered perpendicular magnetic recording media, etc. where conventional vibrating sample magnetometer fails to perform [5, 6]. Research has been carried out to achieve the enhancement of the AHE starting from homogeneous bulk magnetic systems to the ultrathin magnetic heterostructures. Accordingly, the AHE has been studied in various magnetic films such as Fe, Ni and Co single layers and also in various magnetic heterostructures such as Fe/Cr, Fe/Cu, Fe/Gd and Fe/Si multilayers [7-14].

In our earlier studies, we reported large enhancement of AHE in Si/Fe multilayer systems [12, 14, 15]. We observed that the surface and interface scattering play very significant role on the enhancement of the AHE. Recent studies on AHE show significant enhancement in the anomalous Hall signal from magnetic layers sandwiched between oxide layers such as $SiO_2$, MgO, etc. [16-18]. High sensitivity was reported in $SiO_2$/FePt/$SiO_2$ sandwich structure films [16]. Similar high sensitivity was reported by Kopnov et al. in CoFeB film through controllable oxidation [18]. Zhu et al. have studied MgO/CoFeB/Ta/MgO multilayers by introducing MgO as the buffer and capping layer [17]. It is also reported that the interface scattering in metal-insulator nanocomposite films such as $FeSiO_2$, $CoSiO_2$, $NiFeSiO_2$ significantly enhances AHE [2, 3, 19]. Being motivated by the influence of the insulating layers on the enhancement of the AHE in the magnetic layers, in this paper, we have investigated the simplest structure, viz. $SiO_2/Fe/SiO_2$ trilayers.

## 2. Experimental details

Sequential deposition of the $SiO_2$ and Fe layers was carried out onto glass and silicon substrates by dc magnetron sputtering at ambient temperature for obtaining the $SiO_2/Fe/SiO_2$ trilayer structures. The base pressure of the deposition chamber was $2 \times 10^{-6}$ mbar. Using an Fe target, sputtering of the Fe film was carried out at an argon pressure of $4 \times 10^{-3}$ mbar and a DC power of 40 W. For the deposition of the $SiO_2$ layer, reactive sputtering of a Si target was performed in a mixture of Ar and $O_2$ gases at a DC power of 10 W. The flow of $O_2$ was maintained at $0.32 \times 10^{-3}$ mbar (~ 8% of $p_{Ar}$) using a mass flow controller. The stoichiometric composition of the silicon oxide layer as analyzed through X-ray photoelectron spectroscopy (XPS) was $SiO_{1.9}$, which is close to the ideal $SiO_2$. The calibration of the thickness of the layers was done using an Ambios XP2 surface profiler. After making trial runs to ensure the required composition of the $SiO_2$, the trilayers of the form $SiO_2(600 \text{ Å})/Fe(t_{Fe})/SiO_2(400 \text{ Å})$ were grown. Here, $t_{Fe}$ is the nominal Fe layer thickness which is varied from 30 to 300 Å for obtaining different samples. In order to have better uniformity and lesser defects, the thickness of the buffer and capping $SiO_2$ layers were chosen relatively larger i.e. 600 and 400 Å, respectively.

Such thick buffer and capping layers have also been utilized in some other systems [20]. For Hall effect measurements, Cu contact pads of thickness 1000 Å were deposited over the bottom $SiO_2$ (600 Å) layer prior to the deposition of the Fe layer using a shadow mask. The structural properties of the samples were analyzed through X-ray diffraction (XRD) using a Philips PANalytical X'Pert Pro X-ray diffractometer with $CuK_\alpha$ radiation. The microstructural analysis of the samples were done by high resolution transmission electron microscopy (HRTEM) using JEOL JEM 2100F instrument operating at an accelerating voltage of 200 kV. Magnetic measurements of the samples were done at 300 K by a vibrating sample magnetometer, an attachment of a Physical Property Measurement System of Quantum design, Inc. Hall effect measurements were carried out at 300 K by circular four probe method with the perpendicular magnetic field varying up to 27 kOe.

## 3. Results and discussions

*3.1. XPS studies for the composition analysis of the $SiO_2$ and Fe layers*

To verify the stoichiometric composition of the $SiO_2$ layer, X-ray photoelectron spectroscopy (XPS) has been performed on various $SiO_x$ samples prepared under different experimental conditions by varying the $O_2$ pressure and the power applied to the target. The data were analyzed using XPS Peak 4.1 software package. Figure 1(a) shows the Si 2p spectra centered at 103 eV. Similarly, the O 1s spectra centered at 532.7 eV is shown in figure 1(b). From the analysis of these XPS data, the stoichiometric composition of the silicon oxide layer obtained is $SiO_{1.9}$ which is very close to the ideal case, i.e. $SiO_2$. This $SiO_{1.9}$ phase was obtained by preparing the films at a DC power of 10 W and oxygen pressure $P_{O2}$ = 8% of $P_{Ar}$. We have used these sputtering conditions for the $SiO_2$ layers while preparing all the multilayer samples.

To verify the nature of the Fe layer in the trilayer samples, we have taken the XPS measurements on single Fe films grown under similar conditions. Figure 1(c) shows the representative Fe-2p core level XPS spectra of a Fe (30 Å) film. As can be seen from the figure, the Fe-$2p_{3/2}$ and Fe-$2p_{1/2}$ peaks centered around 706.6 and 719.8 eV, respectively. These two peaks are separated by 13.2 eV which is the characteristics of metallic behavior of the Fe film as reported in the literature [21, 22]. Therefore, the Fe layer grown by us is essentially in pure form. However, the AHE data of the samples, as will be discussed in section III-D, shows the absence of the AHE signal for the Fe (30 Å) film sandwiched between two $SiO_2$ layers. This indicates that the $SiO_2$ layers in the $SiO_2$/Fe($t_{Fe}$)/$SiO_2$ trilayer samples oxidize the sandwiched Fe layer and hence resulting in the disappearance of the AHE for $t_{Fe}$ = 30 Å.

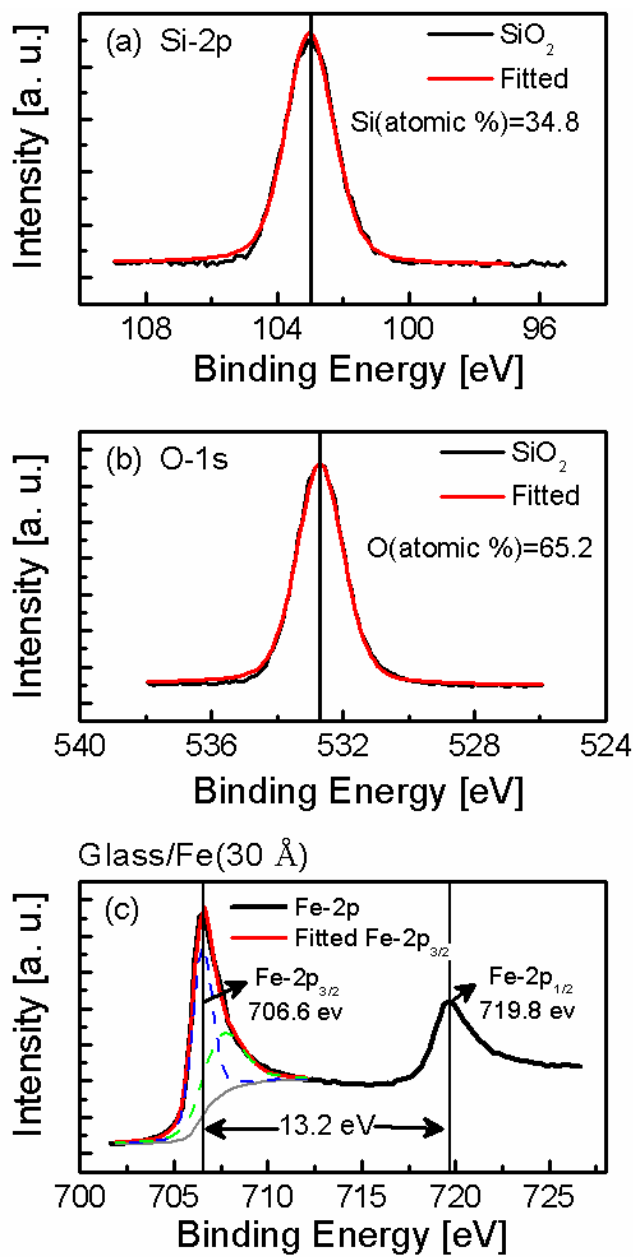

Figure 1 X-ray photoelectron spectra (XPS) of (a) Si 2p (centered at 103 eV) and (b) O 1s (centered at 532.7 eV) for the SiO$_2$ layer. The stoichiometric composition obtained is SiO$_{1.9}$ which is close to the ideal case of SiO$_2$. (c) The Fe-2p core level XPS spectra for the single Fe (30 Å) film indicating that the Fe layer is pure.

*3.2. Structural and microstructural studies*

The thickness of the Fe layer in our $SiO_2/Fe/SiO_2$ sandwich structures ranges from 30 to 300 Å. In the thin film samples of nanoscale thicknesses, the structural, microstructural and topological properties can have a great impact on the electric and transport properties of the materials, especially when the material involved is ferromagnetic and it is sandwiched between two insulating $SiO_2$ layers. The XRD data of all the samples as shown in figure 2(a) indicates that the samples are nanocrystalline with (110) texture. The absence of the Fe(110) peak for $t_{Fe}$ = 30 Å is due to the weak diffraction of the X-rays by the nanosized Fe grains in the samples. An increase in the intensity of the Fe(110) peak with the increase of $t_{Fe}$ as observed from the XRD data of the samples is an indication of the improvement in the crystallinity of the Fe layer. The grain size calculated for the Fe(110) crystallites using Scherrer formula [23] ($\delta = 0.9\lambda/\beta\cos\theta$) increases with the increase of $t_{Fe}$ as shown in figure 2(b). As $t_{Fe}$ decreases from 300 to 50 Å, the grain size of the Fe (110) crystallites in the samples also decreases from about 165 to 50 Å. This decrease of the grain size with the decrease of $t_{Fe}$ increases the surface-to-volume ratio of the grains that in turn increases the complexities in the electrical conduction of carriers at the metal-insulator interface and hence the magnetotransport properties.

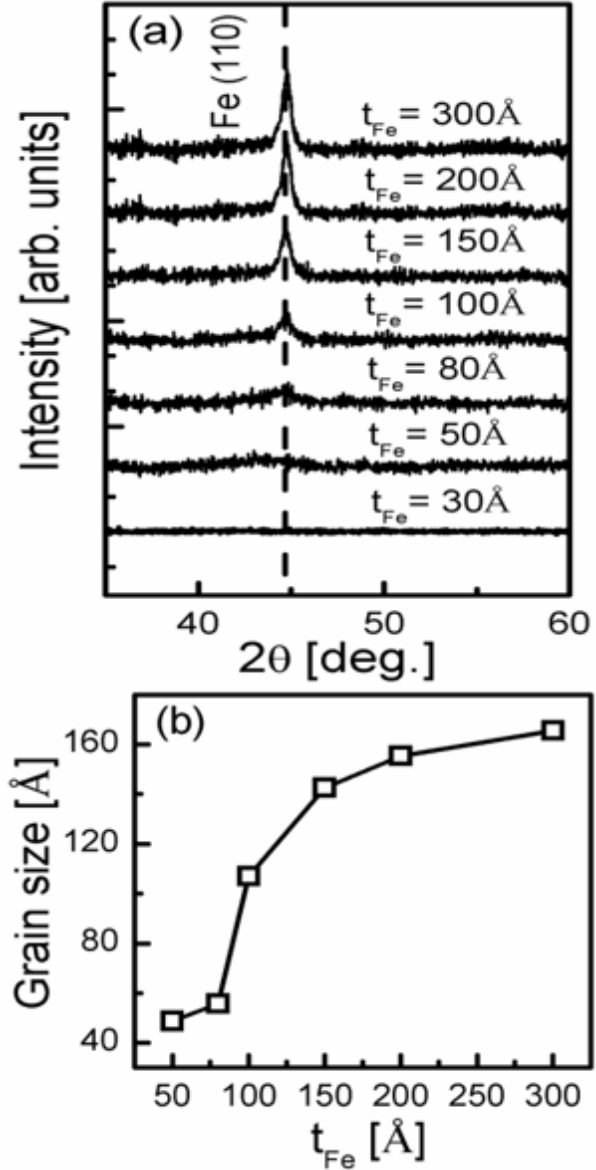

Figure 2 (a) XRD patterns of the $SiO_2$(600 Å)/Fe($t_{Fe}$)/$SiO_2$(400 Å) trilayers. The vertical dashed line indicates the standard bulk $2\theta$ value for the Fe(110). (b) The grain size calculated from the XRD data is plotted as a function of $t_{Fe}$ where the open squares are the data points and the solid line is a guide to the eyes.

As a support to our explanation for the XRD data, the microstructural and morphological analysis of the trilayer samples were done through imaging and selected area electron diffraction (SAED) by high resolution transmission electron microscopy (HRTEM). Figures 3(a) and (b) show the representative HRTEM images of the samples with $t_{Fe}$ = 50 and 100 Å, respectively. As can be seen from the images, the separation between the Fe grains is larger for $t_{Fe}$ = 50 Å than that for $t_{Fe}$ = 100 Å. This suggests that as $t_{Fe}$ decreases the separation between the Fe grains increases, leading to a discontinuous Fe layer. This discontinuity in the Fe layer modifies the interface which in turn can influence the transport properties. The subject of interest here i.e. the AHE that has its origin from a scattering event (skew scattering / side jump) can thus be affected. The SAED pattern of the trilayer with $t_{Fe}$ = 50 Å in the inset of figure 3(a) shows the diffraction rings corresponding to the Fe layer alone. When compared with the XRD data, there are additional Fe(200) and Fe(211) rings observed in the HRTEM diffraction data. The typical grain size obtained from the HRTEM is about 60 Å for this film. This value is slightly larger than that obtained from the XRD data (i.e. 50 Å) of the sample.

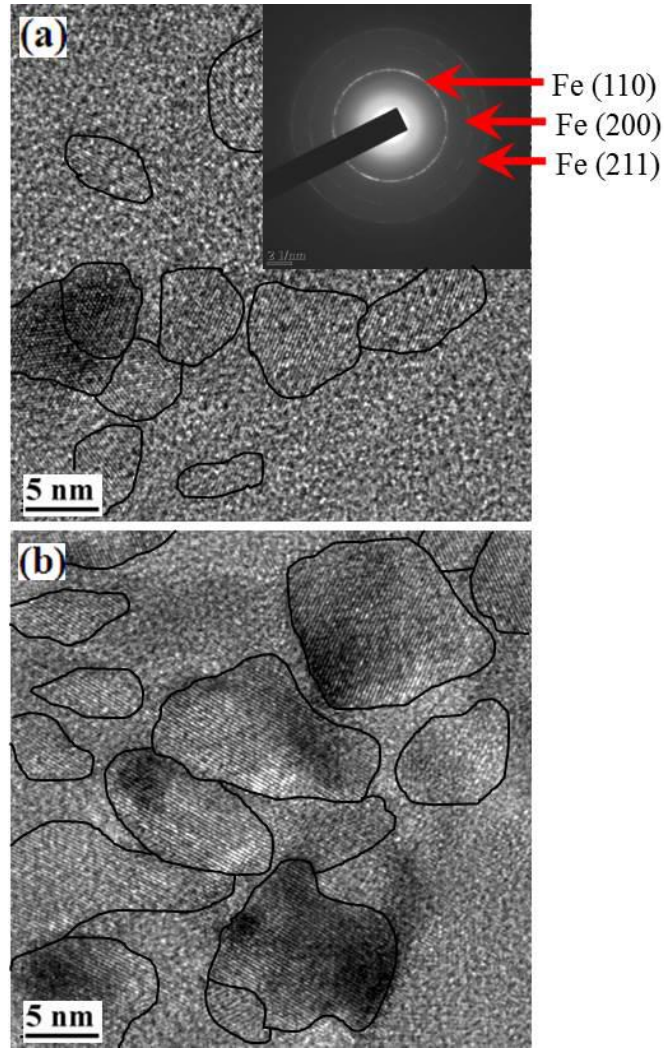

Figure 3 The representative HRTEM images of the SiO$_2$(600 Å)/Fe(t$_{Fe}$)/ SiO$_2$(400 Å) trilayers for (a) t$_{Fe}$ = 50 Å and (b) t$_{Fe}$ = 100 Å. The solid curves drawn represent the boundary of the grains. The grains in (b) are closer to each other as compared to that of (a). The inset of figure (a) shows the selected area electron diffraction pattern for t$_{Fe}$ = 50 Å.

*3.3. Magnetic Properties*

The magnetization measurements on the samples were performed at 300 K by applying magnetic field parallel as well as perpendicular to the film plane up to 50 kOe. Shown in figure 4 are the in-plane and perpendicular MH loops of the samples. For all the samples except for t$_{Fe}$

=30 Å, the magnetization in the in-plane direction saturates easily within a few tens of Oersteds of a magnetic field whereas in the perpendicular direction the magnetization saturates at a relatively higher field of about 2 T. The coercivity $H_c$ obtained from the in-plane magnetization data (< 60 Oe) of the samples is considerably smaller than that obtained from the perpendicular MH data (150 Oe). The remanent magnetization of the samples in the in-plane direction is more than the perpendicular direction. These indicate that the samples are soft ferromagnetic having strong in-plane magnetic anisotropy. The non-saturation of the in-plane and perpendicular MH loops for $t_{Fe}$ = 30 Å is an indication of the presence of superparamagnetic fine Fe grains. Thus, with the decrease of $t_{Fe}$, there is an increase in the fraction of the superparamagnetic fine Fe grains due to the significant decrease in the grain size which is also observed from the XRD and HRTEM data of the samples.

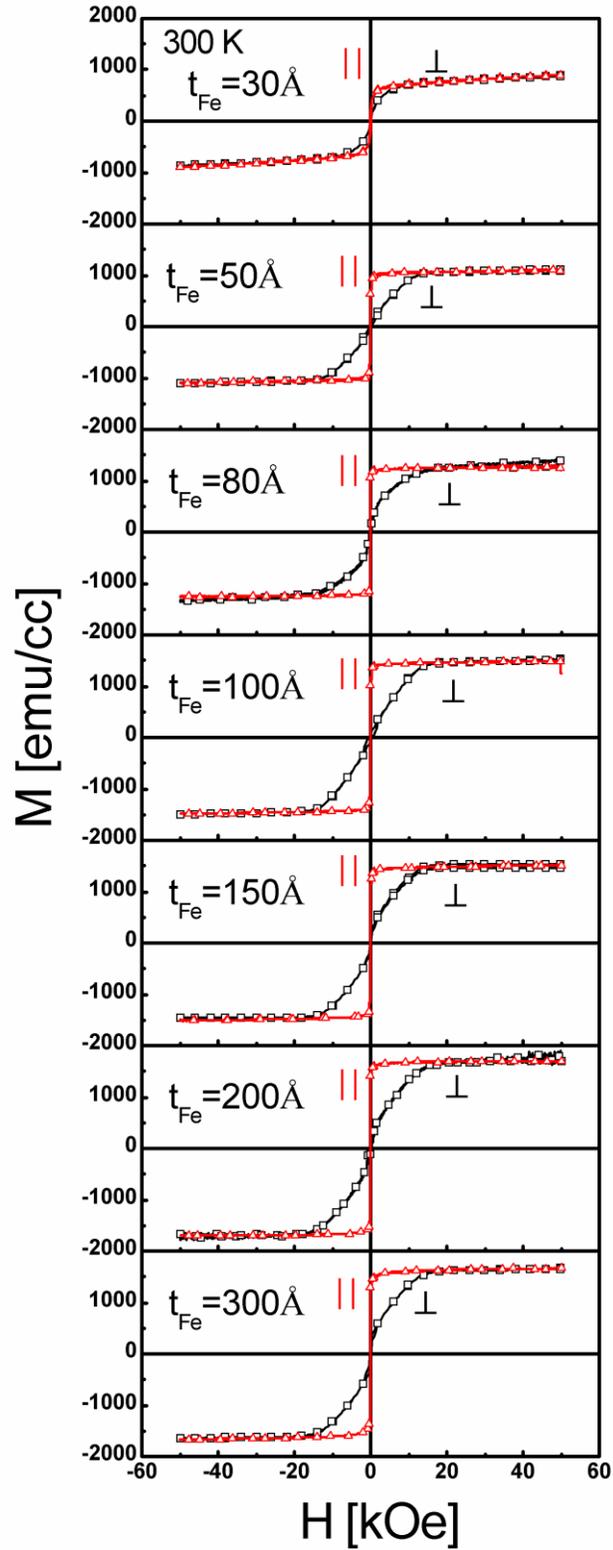

Figure 4 The in-plane (∥) and perpendicular (⊥) MH-loops of the $SiO_2$(600 Å)/Fe($t_{Fe}$)/ $SiO_2$(400 Å) trilayer samples. The non-saturation of the MH-loops is observed for $t_{Fe}$ = 30 Å.

From the analysis of the magnetization data of the samples, the saturation magnetization ($M_s$) obtained is plotted as a function of $t_{Fe}$ as shown in figure 5(a). As $t_{Fe}$ increases, the $M_s$ increases towards the bulk value (1714 emu/cc [24]). This kind of behavior is generally observed in thin films and multilayers. The formation of the magnetically inactive region at the interface is normally observed in the layered systems consisting of ferromagnetic and nonmagnetic materials due to interlayer mixing and solid state alloying reaction between the two materials [22, 25]. Interface oxidation is an additional factor in our samples. The magnetically inactive region formed at each interface can be deduced from the fitting of the $M_s$ data as shown in figure 5(b). This fitting is according to the equation $M_s(t_{Fe}) = M_0(1 - 2\delta/t_{Fe})$, where $M_0$ denotes the bulk magnetization of Fe and $\delta$ is the thickness of the magnetic dead layer at each interface [26, 27]. The deduced value of $\delta$ for the samples is 9.8 Å. This large value is due to the oxidation of the Fe interfaces.

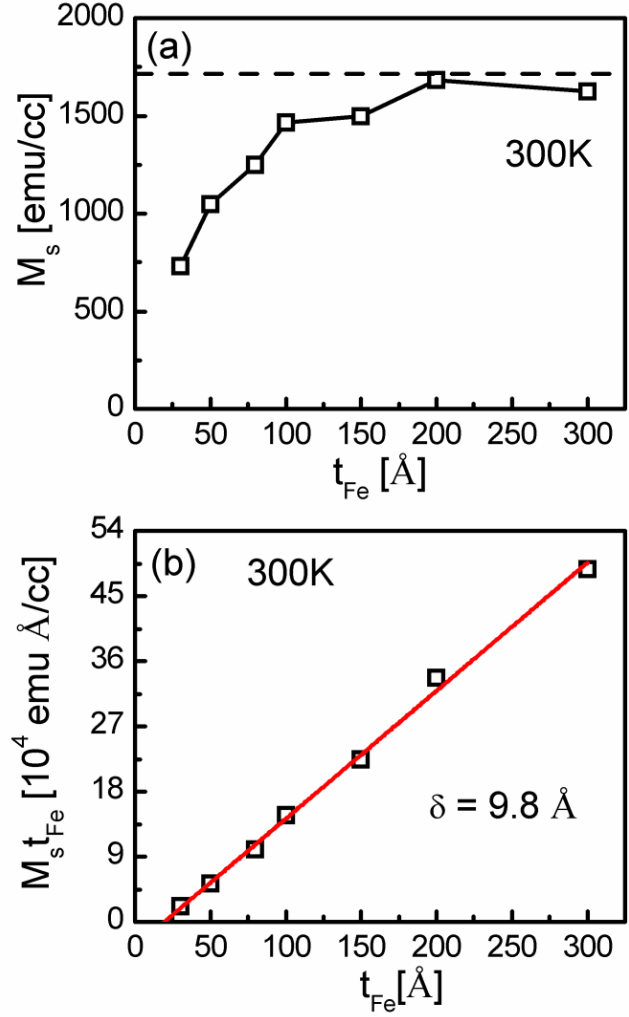

Figure 5 (a) The $t_{Fe}$ dependence of $M_s$. (b) $M_s t_{Fe}$ versus $t_{Fe}$ according to the equation $M_s(t_{Fe}) = M_0(1 - 2\delta/t_{Fe})$, where the open squares are the data points and the solid line is the fitted one.

The temperature dependence of the magnetization of the samples was studied through zero field cooled (ZFC) and field cooled (FC) magnetization measurements. Shown in figure 6 are the ZFC-FC curves of the samples taken in the temperature range of 4.2-350 K in an in-plane magnetic field of 50 Oe. In the ZFC measurement, the sample was first cooled up to 4.2 K, then a small in-plane magnetic field of 50 Oe was applied and the data were taken by increasing the temperature in the range 4.2-350 K. After ZFC measurement, the samples were again cooled to 4.2 K in the presence of the field of 50 Oe. The FC magnetization measurements were taken while warming up the sample from 4.2 to 350 K in the presence of 50 Oe.

In a magnetic specimen, the magnetic anisotropy aligns the spins in a preferred direction. In ZFC measurement, when the sample was cooled through the ordering temperature, the spins get locked in random directions, giving rise to a nearly zero net magnetization. Application of a small field of 50 Oe at the low temperatures energetically favors the reorientation of the moments in the direction of the applied field. This result in the increase of the ZFC magnetization up to a particular temperature called blocking temperature ($T_B$). At temperatures above $T_B$, the thermal energy starts dominating over the anisotropy energy which results in the decrease of the magnetization. The presence of a broad maximum centered at $T_B$ is an indication of the distribution in the volume of the Fe grains, the moments of which progressively undergoes transition from blocked state to unblocked state at different $T_B$ values. The value of $T_B$ obtained from the ZFC magnetization data of the samples varies from 106 K to 311 K as $t_{Fe}$ increases from 30 to 300 Å. Similar to the ZFC magnetization, the FC magnetization of the samples also shows a change with the temperature. As the temperature decreases below $T_B$, the FC magnetization of the samples with $t_{Fe}$ = 200 and 300 Å show a large decrease, indicating that the applied field is not strong enough to align the spins in the field direction. No significant change in the FC magnetization as observed for the samples with $t_{Fe}$ from 30 to 150 Å below $T_B$ suggests that these samples are less anisotropic. This decrease of the magnetic anisotropy with the decrease of $t_{Fe}$ is a consequence of the decrease of the grain size which is also observed from the XRD data of these samples. The observed $T_B$ values in the samples indicates the presence of thermally unstable, low anisotropic superparamagnetic fine Fe grains together with the bigger ferromagnetic Fe grains which lead to the saturation of MH loops (except for $t_{Fe}$ = 30 Å) with nonzero $H_c$ and remanence values.

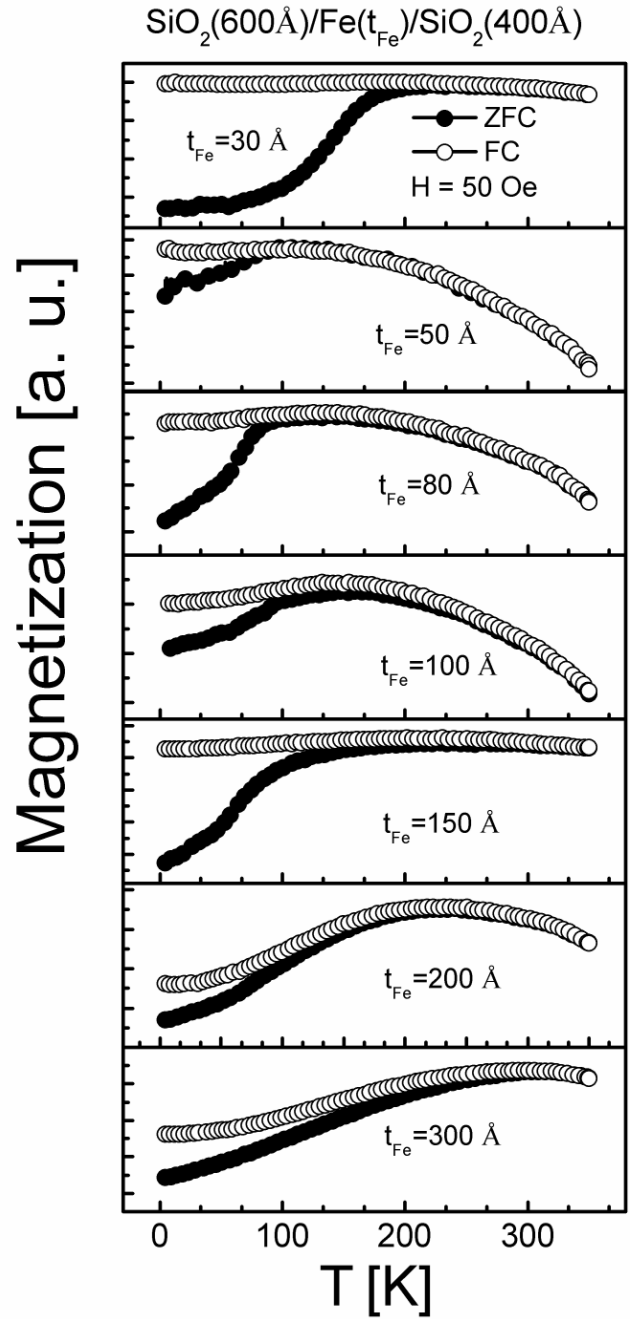

Figure 6 The ZFC-FC curves of the $SiO_2$(600 Å)/Fe($t_{Fe}$)/ $SiO_2$(400 Å) trilayer samples. The solid circles represent the ZFC data and the open circles represent the FC data.

*3.4. Transport properties*

The transport properties of the samples, as studied through Hall effect measurements at 300 K shows anomalous Hall behavior. In the case of magnetic thin films, when magnetic field applied perpendicular to the film plane, the magnetic field induction, $B = H + (1-D)M \cong H$, as the demagnetization factor D=1. Therefore, the Hall resistance ($R_h$) for magnetic thin films can be described by the empirical equation [20, 28-30],

$$R_h = \frac{\rho_h}{t} = \frac{R_0 H + R_s 4\pi M}{t} \qquad (1)$$

where $\rho_h$, $t$, $R_0$, $R_s$ and $M$ are anomalous Hall resistivity, thickness of the magnetic film, ordinary Hall coefficient, anomalous Hall coefficient and perpendicular component of magnetization, respectively. The first term in the equation (1) is the ordinary Hall effect (OHE) term arising because of the Lorentz force on the charge carriers due to the applied external magnetic field. The second term is related to the sample's magnetization and it represents the anomalous Hall effect (AHE) which is believed to be originating due to the break of the right-left symmetry during spin-orbit scattering of electrons. The anomalous term being magnetization dependent saturates after a certain field called perpendicular saturation field ($H_s = 4\pi M_s$). The saturation anomalous Hall resistance ($R_{hs}^A$) can be calculated by subtracting the ordinary term from the linear fit of the high field regime Hall data. The values of $R_0$, $R_s$ and $H_s$ can be calculated by following the procedure described in refs. 9 and 29.

The $R_h$ of the trilayer samples with different $t_{Fe}$ at 300 K is plotted as a function of H as illustrated in figure 7. In the low field regime, the $R_h$ has a strong linear dependence with the field due to the dominance of the AHE over the OHE. The steep increase of $R_h$ with H can be ascribed to the reorientation of magnetic domains in the applied field direction. After a certain field $H_s$, as the magnetization gets saturated so does the AHE. As a result, the $R_h$ now increases very slowly with H due to the OHE. Two distinct regions of slope observed in the $R_h$ versus H graph clearly indicate that the AHE in these materials is much stronger than the OHE. It can also be seen from the figure that a large increase in the AHE is observed upon decreasing $t_{Fe}$. The AHE signal disappears upon decreasing $t_{Fe}$ below 50 Å. This is because of the loss of ferromagnetism due to the oxidation of the interfaces of the Fe layer which is in contact with the $SiO_2$ layer. This behavior is consistent with the magnetization data of the sample as shown in figure 5(a).

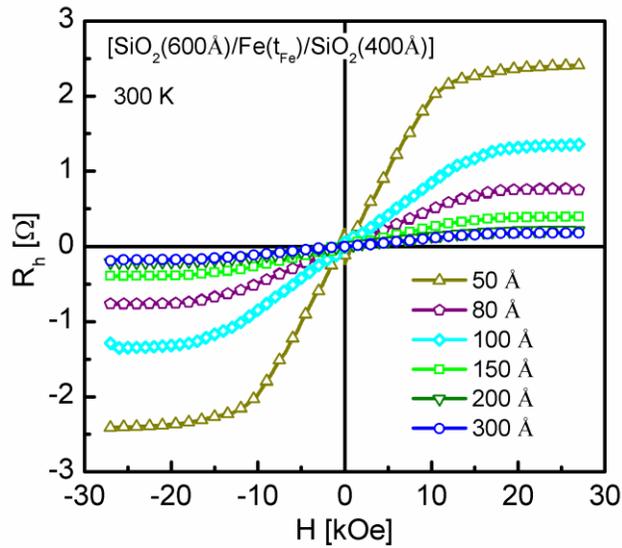

Figure 7 The $R_h$ versus H graph for $SiO_2$(600 Å)/Fe($t_{Fe}$)/ $SiO_2$(400 Å) trilayers with different $t_{Fe}$ at 300 K.

From the analysis of the AHE data, the normalized $R_{hs}^A$ and $R_s$ with respect to that for $t_{Fe}$=300 Å were obtained and they are plotted as a function of $t_{Fe}$ as shown in figure 8. Both $R_{hs}^A$ and $R_s$ show an increasing trend with the decrease of $t_{Fe}$. An increase of about 14 times in $R_{hs}^A$ and about 5 times in $R_s$ has been observed upon decreasing $t_{Fe}$ from 300 to 50 Å. About 3-4 times increase of the AHE signal have already been reported in pure Fe films within the thickness range from 1700 to 80 Å [7]. In our trilayer samples, the largest value of $R_{hs}^{A}$ = 2.3 Ω observed for $t_{Fe}$ = 50 Å is about 4 orders of magnitude larger than that reported for bulk Fe i.e. 0.0002 Ω [1]. Such giant increase in the $R_{hs}^{A}$ as compared with the bulk material is often referred as giant Hall effect (GHE) by some authors [1, 2, 31]. For comparison, we have also performed Hall effect measurements on single Fe films grown by us. Considerable enhancement in the AHE has been observed in the Fe layer which is in sandwiched form when compared with that of the single Fe films. A significant increase of about 51% in $R_{hs}^{A}$ of a Fe(50 Å) film has been observed upon sandwiching it between two $SiO_2$ layers. This shows the importance of insulating $SiO_2$ layer as buffer and capping layers. This large increase in the AHE signal can be attributed to the increase of the electron scattering due to the metal-insulator interface. A change in the size effects, impurity contents and the defects due to the decreased $t_{Fe}$ increases the scattering rate of the charge carriers, thus resulting in the enhancement of the AHE. Our earlier study on the AHE in Si/Fe multilayers shows a large enhancement of about 60 times in $R_{hs}^{A}$ and 80 times in $R_s$ upon decreasing the Fe layer thickness from 100 to 20 Å [12]. The ferromagnetic metal-semiconducting interface plays a significant role in such large enhancements of the AHE in the Si/Fe multilayers. The enhancement observed in the case of the present $SiO_2$/Fe/$SiO_2$ trilayers is quite small when compared with that of the Si/Fe multilayers. This is because the Fe layer being in contact with the two $SiO_2$ layers on the top and bottom gets partially oxidized by receiving oxygen atoms from the $SiO_2$. This oxidation of the Fe layer further leads to the loss of its ferromagnetism that in turn results in the decrease of AHE. Here, we performed the Hall measurements on the $SiO_2$/Fe($t_{Fe}$)/$SiO_2$ trilayers upon decreasing $t_{Fe}$ from 300 Å down to 50 Å and obtained the largest AHE for $t_{Fe}$ = 50 Å. For $t_{Fe}$ below 50 Å, the large oxidation of the Fe layer due to the SiO$_2$ layers in contact results in the disappearance of the AHE.

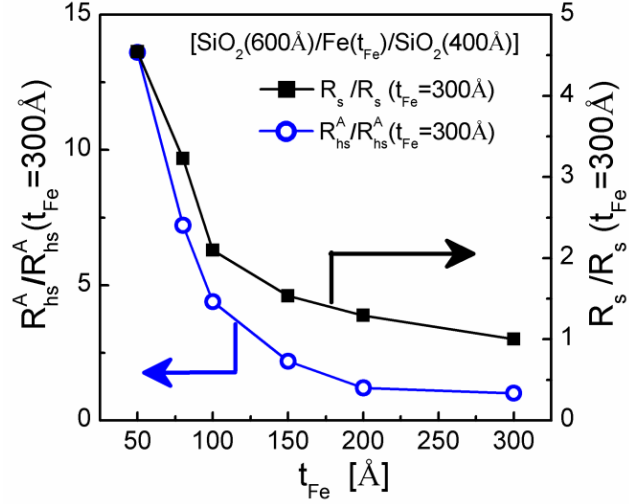

Figure 8 The $t_{Fe}$ dependence of $R_{hs}^A$ and $R_s$ of the SiO$_2$/Fe($t_{Fe}$)/SiO$_2$ trilayer films normalized with respect to that for $t_{Fe}$ = 300 Å. Enhancements of about 14 times in $R_{hs}^A$ and about 4 times in $R_s$ have been observed.

The ordinary Hall coefficient ($R_0$) as extracted from the linear fit of the high field regime Hall effect data of the trilayer samples is normalized with respect to that of $t_{Fe}$ = 300 Å and plotted as a function of $t_{Fe}$ as shown in figure 9. The $R_0$ shows an enhancement of about 7 times upon decreasing $t_{Fe}$ from 300 to 50 Å. The largest value of $R_0$ i.e. 9.5 × 10$^{-10}$ Ω m/T observed in the case of $t_{Fe}$ = 50 Å is about two order of magnitude larger than that of bulk Fe (~ 10$^{-11}$ Ω m/T) [1, 29]. Similar to the case of bulk Fe [29], the sign of $R_0$ for all samples is found to be positive indicating that the conduction mechanism is due to the positive charge carriers or holes. The carrier concentration ($n'$) was calculated using the relation $R_0 = 1/n'e$ where e = 1.6 × 10$^{-19}$ C.

The $n'$ is of the order of $10^{22}$ cm$^{-3}$ for all the samples except for $t_{Fe}$ = 50 Å in which the $n'$ is of the order of $10^{21}$ cm$^{-3}$. The $n'$ as reported for bulk Fe is of the order of $10^{23}$ cm$^{-3}$ [2, 20]. This decrease in $n'$ and hence the increase in $R_0$ of the trilayer films is due to the oxidation of the interfaces of the Fe layer which is in contact with the SiO$_2$ layer.

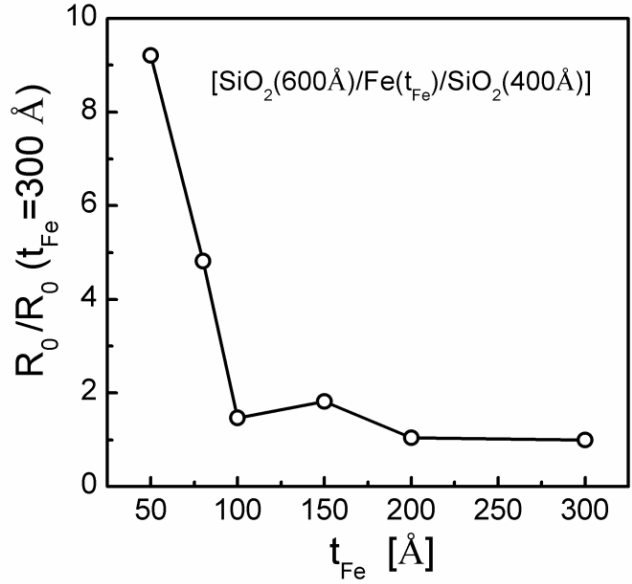

Figure 9 The $t_{Fe}$ dependence of $R_0$ normalized with respect to that of $t_{Fe}$ = 300 Å. The open circles are the data points and the solid line is a guide to the eyes.

The AHE of a ferromagnetic material is the result of what actually occurs microscopically in each domain due to their spontaneous magnetization. To understand how actually the AHE occurs, various mechanisms such as Karplus-Luttinger mechanism [32], skew scattering mechanism [33] and side jump mechanism [34] have been proposed. All these three mechanisms of the AHE suggest a scaling law between the anomalous Hall resistivity ($\rho_h$) and longitudinal electronic resistivity ($\rho$) by considering the scattering of the carriers as the origin. The scaling law between these two can be written as

$$R_s \propto \rho^n \, (\text{or } \rho_h \propto \rho^n),  \qquad (2)$$

where n = 1 and 2 represent the skew scattering and side jump mechanisms of the AHE, respectively. The value of 'n' obtained from the fitting of the experimental data using equation (2) will give the dominant mechanism of the AHE in the samples.

For scaling of the AHE with the resistivity, we have measured the $\rho$ of the trilayer samples at 300 K. Here $\rho$ was measured at H = 0 and it was varied from $4.3 \times 10^{-6}$ Ω m for $t_{Fe}$ = 50 Å to $2.0 \times 10^{-6}$ Ω m for $t_{Fe}$ = 300 Å. The $\rho$ value for pure Fe as reported in the literature is $10.1 \times 10^{-8}$ Ω m [7, 29]. To verify that the enhancement of resistivity is due to the metal-insulator interface, we have compared the resistivity data of our $SiO_2$/Fe(50 Å)/$SiO_2$ trilayer with that of the single Fe (50 Å) film and Si(50 Å)/Fe(50 Å) bilayer sample prepared under similar conditions. The $\rho$ value for Fe (50 Å) film was $4.4 \times 10^{-7}$ Ω m and that for Si(50 Å)/Fe(50 Å) bilayer was $9.6 \times 10^{-7}$. Upon comparing with the $\rho$ of pure Fe we have observed that the enhancement in $\rho$ was quite high (about 43 times) for the $SiO_2$/Fe(50 Å)/$SiO_2$ trilayer in comparison with that of the Fe(50 Å) film (~ 4 times) and the Si(50 Å)/Fe(50 Å) bilayer sample(~ 10 times). This suggests that the enhancement in the resistivity and hence in the AHE of our trilayer samples is mainly due to the Fe-$SiO_2$ interface. Furthermore, the XPS data shown in figure 1(c) also supports the fact that the Fe-$SiO_2$ interface is responsible for the enhanced resistivity.

To understand the mechanism of the AHE in the $SiO_2$/Fe/$SiO_2$ trilayers, we have verified the scaling law between $R_s$ and $\rho$ as given by equation (2). The value of the exponent 'n' can be obtained from the linear fit of the graph between ln $R_s$ and ln $\rho$ as shown in figure 10. The 'n' obtained from the above fitting is 1.9 which is close to 2, thus suggesting side jump as the dominant mechanism of the AHE. A value of n = 1.6 has been reported in the case of epitaxial

Fe films by Sangiao et al. [35]. The side jump as the dominant mechanism of the AHE has also been reported in the case of the Fe films by some authors [36-38]. Therefore, additional sources of scattering due to the surface modifications by the two insulating $SiO_2$ layers at the top and bottom of the Fe layer in our trilayer samples does not change the mechanism of the AHE. This is consistent with the results already reported by Gerber *et al.* [39]. In the case of granular $FeSiO_2$ films, n = 0.5 was reported by Aronzon *et al.* where they claimed the difference in n other than 1 (skew scattering) or 2 (side jump) is due to the tunneling type carrier conduction [2]. The n = 2.14 has also been reported in the case of $Fe_{100-x}(SiO_2)_x$ [40]. These very different values of 'n' in the case of the metal-insulator heterostructure system by different groups demands further theoretical investigation in this field.

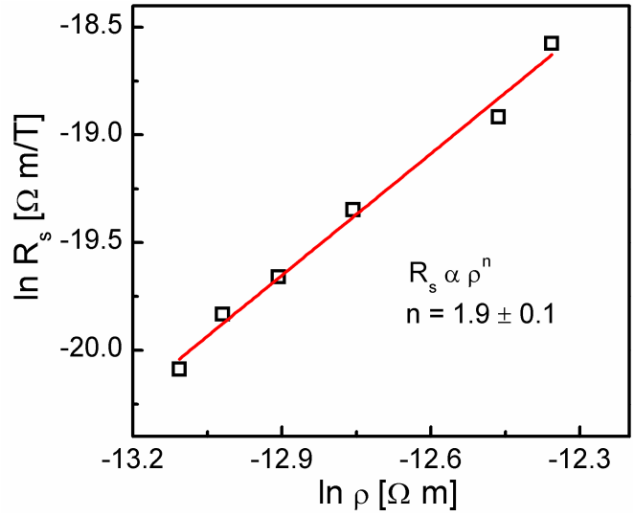

Figure 10 The ln $R_s$ versus ln $\rho$ plot to verify the scaling law between $R_s$ and $\rho$ for the $SiO_2/Fe/SiO_2$ sandwich structured films. Here $\rho$ was measured at H = 0. The open squares are the experimental data points and the solid line is a linear fit according to equation (2), i.e. $R_s \propto \rho^n$.

The anomalous Hall sensitivity S defined by $dR_h/dH$ in the low field linear regime Hall effect data of the samples is plotted as a function of $t_{Fe}$ in figure 11. A large increase of S from 0.1 Ω/T to 2.3 Ω/T was observed upon decreasing $t_{Fe}$ from 300 to 50 Å. Also shown in figure 11 is the $t_{Fe}$ dependence of $H_s$. A decrease of $H_s$ from about 16 kOe to 11 kOe has been observed upon decreasing $t_{Fe}$ from 300 to 50 Å. The S which has a reciprocal relation with the $H_s$ can thus be strongly influenced by this decrease of $H_s$ at lower $t_{Fe}$ [41]. A maximum enhancement of about 156% in S has been observed for Fe(50 Å) film upon sandwiching it between the two $SiO_2$ layers. This enhancement of S in the case of the $SiO_2/Fe/SiO_2$ sandwich films in comparison with the Fe films grown directly on the substrates indicates the role of the metal-insulator interface. When compared with the Fe(50 Å) single film, the maximum decrease of about 90% in $H_s$ has been observed for sandwiched Fe(50 Å) film. This decrease of $H_s$ together with the increased $R_{hs}^A$ due to the metal-insulator interface scattering results in such large enhancement in the S of the $SiO_2/Fe/SiO_2$ sandwich structured films. Thus, the enhancement of the AHE observed in the $SiO_2/Fe/SiO_2$ trilayers envisages us that the material could be a possible candidate for Hall element in the field of magnetic sensors if further improvement in the optimization of the layer thicknesses is achieved.

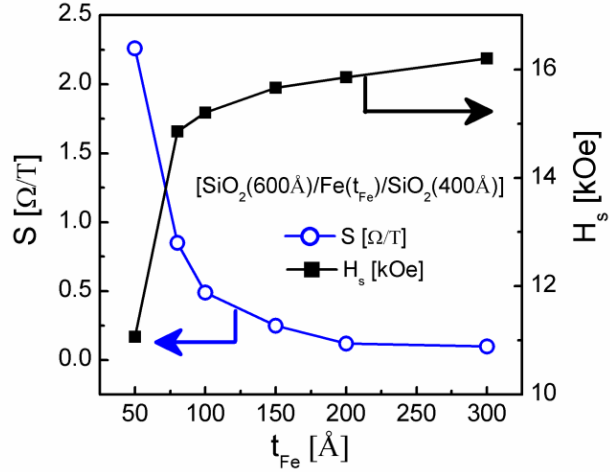

Figure 11 The $t_{Fe}$ dependence of the Hall sensitivity (S) and perpendicular saturation field ($H_s$) of the $SiO_2/Fe(t_{Fe})/SiO_2$ sandwich films.

## 4. Conclusion

In conclusion, we have investigated the structure, microstructure, magnetic and magnetotransport properties of the sputtered $SiO_2/Fe(t_{Fe})/SiO_2$ sandwich structured films at 300 K. As observed from the XRD data and further confirmed from the HRTEM data, the samples have grown in nanocrystalline form with (110) texture. Magnetization measurements both in in-plane and perpendicular direction showed that the samples have in-plane easy axis of magnetization. Both the MH data and ZFC-FC data of the samples show complex behavior with the coexistence of both ferromagnetic and superparamagnetic properties. For the film with $t_{Fe}$ = 30 Å, the non-saturation of magnetization indicates the presence of superparamagnetic Fe grains. The AHE signal has also not been observed. These are because of the loss of ferromagnetism due to the oxidation of the Fe layer which is in contact with the $SiO_2$ layers. For all other samples, the Hall effect measurements suggests a significant enhancement in the AHE with decreasing $t_{Fe}$.

The largest value of $R_{hs}^{A}$ = 2.3 Ω observed for $t_{Fe}$ = 50 Å is about four orders of magnitude larger than that of bulk Fe. Scaling law suggests the side jump as the dominant mechanism behind the observed AHE in the samples. A maximum enhancement of about 156% in S has been achieved by sandwiching the Fe layer between the two $SiO_2$ layers. Our data show that the sensitivity of the trilayer films can further be enhanced if more improvement in the perpendicular anisotropy is achieved by optimizing the Fe layer thickness with the help of more advanced fabrication technologies.